\documentclass[twocolumn,amsmath,amssymb,pra,superscriptaddress]{revtex4}

\usepackage{graphics}
\usepackage{epsfig}
\usepackage{amsmath}
\usepackage{amsfonts}
\usepackage{color}
\usepackage{dcolumn}
\usepackage{bm}
\usepackage{mathrsfs}
\usepackage{float}

\begin{document}

  \bibliographystyle{apsrev}
  
\title{Pseudopotential for Many-Electron Atoms}

      \author{Eric Ouma Jobunga}
       \affiliation{\it  Department  of Mathematics and Physics , Technical University   of Mombasa,\\ P. O. Box 90420-80100, Mombasa, Kenya}
%
%


\begin{abstract}
Electron-electron correlation forms the basis of difficulties encountered in multi-electron systems. Accurate treatment of the correlation problem is likely to unravel some nice physical properties of matter embedded in this correlation. In an effort to tackle this multi-electron problem, two complementary parameter-free pseudopotentials for $n$-electron atoms are suggested in this study. Using one of the pseudopotentials, near-exact values of the groundstate ionization potentials of helium, lithium, and berrylium atoms have been calculated. The other pseudopotential also proves to be capable of yielding reasonable and reliable ionization potentials within the non-relativistic quantum mechanics framework.

\end{abstract}

\maketitle
\section{Introduction}

The theory of quantum many-body systems\cite{Hugenholtz1965} is an effective theoretical structure and solvable approach of understanding the collective behaviour of the interacting many-particle systems. The solution of the many-electron problem is important because electrons determine the physical properties of materials and molecules. Many-body physics is heavily applicable in condensed matter, Bose-Einstein Condensation (BEC) and superfluidity, quantum chemistry, atomic, molecular, nuclear physics, as well as quantum chromodynamics.

Electron correlation energy, among the interacting many-body particles, is defined as the difference between the exact non-relativistic energy eigenvalue of the electronic Schr\"odinger equation and the energy of the single configuration state function (CSF) approximation, commonly called the Hartree-Fock energy \cite{Verdebout2013}. 

Accurate description of electron-electron interaction remains a major challenge in atomic structure calculations \cite{Verdebout2013}. To meet this challenge, a number of different methods have been developed such as the many-body perturbation theory (MBPT) \cite{Tobocman1957}, configuration interaction (CI)\cite{Cremer2013},  density functional theory \cite{Kohn1965}, coupled cluster theories, and different kinds of variational methods \cite{Cramer2002}. Hylleraas-type calculations \cite{Hylleraas1929} is an example of the variational methods in which the interelectronic distance $r_{12}$ is employed explicitly in the construction of the wavefunction resulting into the most accurate eigenvalues, although computationally expensive.

A pseudopotential is an effective potential used as an approximation for the simplified description of complex atoms, molecules, and other quantum systems. The use of pseudoptentials was first introduced by Fermi \cite{Cohen1984}. Hellmann \cite{Hellmann1935} subsequently developed a pseudopotential model for atoms which has been extensively used in atomic scattering \cite{Callaway1969}. The use of pseudopotential method in the many-body problems is computationally less expensive and has the potential of revealing the underlying processes in the interaction dynamics.

In this work, a central screening potential in an independent particle model introduced in our previous papers \cite{Jobunga2017, Jobunga2017b}, is extended to an $n-$electron atom. The generalised parameter-free pseudopotentials developed in this work are used to evaluate the groundstate ionization potentials of atoms with upto $12$ electrons. Also, the eigenvalues of some of the excited states of lithium atom have been calculated.

\section{Theory}
The non-relativistic Hamiltonian of an $n$-electron system with a nuclear charge $Z$ is given by

    \begin{equation}
    \mathrm{H} = \sum_i^n \left[\frac{p_i^2}{2} - \frac{Z}{r_i} +  \sum_{j\neq i}^{n-1}\frac{1}{|\mathbf{r}_i-\mathbf{r}_j|} \right]
    \end{equation}
where the first term on the right corresponds to the kinetic energy of the $i^{\mathrm{th}}-$electron, the second term corresponds to the  interaction of the $i^{\mathrm{th}}-$electron with the nuclear charge, and the last term in the summation corresponds to the interaction between the $i^{\mathrm{th}}-$ and $j^{\mathrm{th}}-$ electron. The second and the last term form the potential energy function of a bound $n$-electron system.

 In our previous work \cite{Jobunga2017}, it was shown that the electron correlation interaction simplifies to
 \begin{equation}
 \frac{1}{|\mathbf{r}_i-\mathbf{r}_j|} = \frac{1}{\sqrt{r_i^2 + r_j^2}} \label{eq:0}
 \end{equation}
 using the lowest-order multipole expansion approximation. In the independent particle approximation method, the potential function 
 \begin{equation}
 V(r_i, r_j) = -\frac{Z}{r_i} + \sum_{j\neq i}^{n-1} \alpha_{ij}\,\frac{1}{\sqrt{r_i^2 + r_j^2}}  \label{eq:pt1}
 \end{equation}
for the $i^{\mathrm{th}}-$electron of the system. The coefficient $\alpha_{ij}$ defines the ratio for partioning the correlation energy. Conventionally, factor $1/2$ which assumes equal sharing of the correlation energy between the interacting electrons is usually preferred. The interaction potential $ V(r_i, r_j)$ can be completely separated, by minimizing it with respect to the spatial co-ordinates, as
\begin{equation}
 V(r_i) = -\frac{Z}{r_i} + \frac{(n-1)}{2}\,\frac{\left[\frac{2Z}{(n-1)}\right]^{1/3}}{r_i} \label{eq:pt1a}
 \end{equation}
  or
\begin{equation}
 V(r_i) = -\frac{Z}{r_i} + (n-1)\,\frac{\left[\frac{Z\, f(r_i,r_j)}{2(n-1)}\right]^{3/5}}{r_i} \label{eq:pt1b}
 \end{equation}
  where the correlated two-dimensional function 
 \begin{equation}
 f(r_i, r_j) = \frac{r_i^2}{r_i^2 + r_j^2} \label{eq:2}
 \end{equation}
 is equivalent to the partition function introduced in ref.\cite{Jobunga2017b}. The exact evaluation of the electron correlation term of equation (\ref{eq:0}), resulting into equation (\ref{eq:pt1b}), is shown in appendix \ref{app:A}. As a limitation to the corresponding pseudopotential, the value of the function ${[f(r_i,r_j)]^{\frac{3}{5}}}$ in equation (\ref{eq:pt1b}) cannot be evaluated exactly but can only be approximated by taking its expectation value relative to the some trial wavefunction of the $j^{\mathrm{th}}$ electron. Equation (\ref{eq:pt1a}) is derived using the equal partitioning of the correlation energy between the interacting electrons \cite{Jobunga2017} while equation (\ref{eq:pt1b}) is derived using the exact partition function introduced in our previous work \cite{Jobunga2017b}.
 
In our working, the expectation value of the correlated function given by equation (\ref{eq:2}), expressed in terms of one of the radial co-ordinate, is evaluated approximately as
\begin{equation}
 \langle [f(r_i, r_j)]^{\frac{3}{5}} \rangle \approx 1 - \left[ \frac{27}{25} + \frac{6}{5}Zr_i - \frac{6}{125\,Zr_i}\right]\, \exp (-2Zr_i). \label{eq:pt3}
 \end{equation}
  Appendix \ref{app:B}  shows the explicit method used in arriving at this expectation value.
The approximation in equation (\ref{eq:pt3}) if employed in the independent electron potential, defined in equation (\ref{eq:pt1b}), is found to be of a better agreement with the experimental results. However, more work needs to be done in order to refine this approximation further. 

From equation (\ref{eq:pt1b}), the effective interacting potential for a single electron becomes
 \begin{equation}
 V(r_i) = -\frac{Z}{r_i} + (n-1)\,\frac{\left[\frac{Z}{2\,(n-1)}\right]^{\frac{3}{5}}\,\zeta(r_i)}{r_i} \label{eq:pt5}
 \end{equation}

 with $\zeta(r_i)$ is the expectation value of the correlated function given by equations (\ref{eq:pt3}). It may be important to note equation (\ref{eq:pt5}) is consistent with the form of a pseudopotential prescribed by Hellmann \cite{Hellmann1935}. With the pseudopotentials formulated, the one electron Hamiltonian 
 \begin{equation}
 h(r_i)= \frac{p_i^2}{2} + V(r_i) \label{eq:pt6}
 \end{equation}
 
 is defined. It is evident that the first term of the single electron potential defined in equation (\ref{eq:pt5}) is the electron-nuclear interaction, and the second term yields the screening potential of the active electron from the other electrons. In equation (\ref{eq:pt1a}), the charge screening parameter is constant where as in equation (\ref{eq:pt5}), the charge screening parameter is $r$-dependent.  The eigenvalues of the $n$-electron system can then be evaluated as
 \begin{equation}
   E_{\alpha_1, \alpha_2\, \cdots}   = \sum_i^n\langle \Psi(\mathbf{r}_1,\mathbf{r}_2,\cdots)|h_i(r_i)|\Psi(\mathbf{r}_1,\mathbf{r}_2,\cdots) \rangle \label{eq:pt7}
 \end{equation}
 where $\Psi(\mathbf{r}_1,\mathbf{r}_2,\cdots)$ is the total wavefunction expanded as a Slater determinant and $\{\alpha_1, \alpha_2,  \cdots\}$ are the set of quantum numbers representing the configuration of the quantum state for the system. As an example, equation (\ref{eq:pt7}) is evaluated further using lithium atom, consisting of three electrons, as a case study.

The total wavefunction for lithium atom can be expressed as
\begin{equation}
\Psi(\mathbf{r}_1,\mathbf{r}_2,\mathbf{r}_3) = \frac{1}{\sqrt{3}}\sum_{i=1}^3 [1+(-1)^P\mathrm{P}_{jk}]\phi_{\alpha_i}(\mathbf{r}_i)\,\phi_{\alpha_j}(\mathbf{r}_j)\,\phi_{\alpha_k}(\mathbf{r}_k)  \label{eq:pt8}
\end{equation}
 where $\mathrm{P}_{jk}$ is the permutation operator which interchanges the positions of $j^{\mathrm{th}}-$ and $k^{\mathrm{th}}-$ electron, and $P=0\; \mathrm{ or }\; 1$ for  a symmetric or anti-symmetric wavefunction respectively.
 
 Substituting equation (\ref{eq:pt8}) into equation (\ref{eq:pt7}) and simplifying yields
 \begin{equation}
 \begin{split}
 \epsilon_{\alpha_i}  =& \frac{1}{3}\sum_{i=1}^3 \langle \phi_{\alpha_i}(\mathbf{r}_i)|h(r_i)|\phi_{\alpha_i}(\mathbf{r}_i)\rangle\\
  &+\langle \phi_{\alpha_i}(\mathbf{r}_i)|h(r_i)|\phi_{\alpha_i}(\mathbf{r}_i)\rangle\, \delta_{\alpha_j \alpha_k}\\ 
  &+\langle \phi_{\alpha_i}(\mathbf{r}_i)|h(r_i)|\phi_{\alpha_j}(\mathbf{r}_i)\rangle\, \delta_{\alpha_i \alpha_j}\\ 
  &+ \langle \phi_{\alpha_i}(\mathbf{r}_i)|h(r_i)|\phi_{\alpha_k}(\mathbf{r}_i)\rangle\, \delta_{\alpha_i \alpha_k}\\
  &+ \langle \phi_{\alpha_i}(\mathbf{r}_i)|h(r_i)|\phi_{\alpha_k}(\mathbf{r}_i)\rangle\, \delta_{\alpha_i \alpha_j}\delta_{\alpha_j \alpha_k}\\ 
  &+ \langle \phi_{\alpha_i}(\mathbf{r}_i)|h(r_i)|\phi_{\alpha_j}(\mathbf{r}_i)\rangle\, \delta_{\alpha_i \alpha_k} \delta_{\alpha_j \alpha_k} \\
  \end{split}
 \end{equation}
 where $\delta_{\alpha_i \alpha_j}$ is the Kronecker delta whose value is $1$ if ${\alpha_i= \alpha_j}$ and $0$ otherwise.
 
 The eigenvalue $\epsilon_{\alpha_i}$ corresponding to state $\alpha_i$ for an $n$-electron atom can be generalized as
\begin{equation}
   \epsilon_{\alpha_i} = \frac{m}{n}\, \langle \phi_{\alpha_i}| h(\mathbf{r}_i)| \phi_{\alpha_i} \rangle \label{eq:pt9}
\end{equation} 
where $m$ refers to the number of non-vanishing integrals out of the possible $n!$ permutations. It is important to note that out of these integrals, there is only $1$ direct integral and the remaining $m-1$ are exchange integrals. For lithium, ${m/n=2/3}$. In principle, the integer $m$ can be determined from the groundstate configuration of the atom.   
 
\section{Results and Discussions}

We have developed two pseudopotentials for an $n$-electron system given by equations (\ref{eq:pt1a}) and (\ref{eq:pt1b}). The pseudopotentials are used to calculate the groundstate ionization potentials for $n$-electron atoms as shown in table \ref{tab1} with ${2\leq n \leq 12}$. Our results are compared with available reference data \cite{amo:Bransden1990}. In generating our results, a B-spline radial box of $600$ B-splines, ${r_{\mathrm{max}}=200}$, $k=10$, and a non-linear knot sequence is used.
\begin{table}[!ht]
    \centering
    \begin{tabular}{cccccc}
    \hline
    $n$ & Atom& $m/n$ & Present$_1$& Present$_2$& Ref.(eV)  \\
   \hline
   \hline
       $2$& He&2/2   & 24.76   & 35.21  & 24.60   \\                                                     
       $3$& Li&2/3   & 5.50    &  4.97  &5.39   \\                          
       $4$& Be&3/4   & 9.40    & 8.90   &9.32  \\  
       $5$& B &3/5   & 10.65   & 8.08   &8.30   \\ 
       $6$& C &4/6   & 15.95   & 12.29  &11.26 \\                                  
       $7$& N &4/7   & 17.73   & 13.82  & 14.53  \\                                                     
       $8$& O &4/8   & 19.53   & 15.37  & 13.62  \\                          
       $9$& F &4/9   & 21.34   & 16.92  & 17.42  \\  
       $10$&Ne& 5/10 & 28.96   & 23.11  &21.56  \\ 
       $11$&Na& 2/11 & 5.55    & 5.30   &5.14  \\                                  
       $12$&Mg& 3/12 & 8.94    & 8.58   &7.65  \\                                                                                             
    \hline
    \end{tabular}
    \caption{Some numerically calculated ionization potentials for $n$-electron atoms using the present pseudopotentials versus the reference values \cite{amo:Bransden1990}.  Present$_{1}$ and Present$_2$ are the results given by the Hamiltonians with equations (\ref{eq:pt1a}) and (\ref{eq:pt5}) as the pseudopotentials respectively. The results presented are truncated to $2$ d.p. }
    \label{tab1}
  \end{table}
The groundstate ionization potentials calculated using the pseudopotential in equation (\ref{eq:pt1a}) are in very good agreement with the reference values only for spherically symmetric cases of helium, lithium, and berrylium atoms. The results are therefore only reliable for these three atoms. Following the discussion for helium atom in reference \cite{Jobunga2017}, the discrepancy between groundstate ionization potentials for the three atoms may be attributed to the relativistic and other higher-order corrections in the interaction Hamiltonian.  For atoms with non-$s$ states in the groundstate configuration, the role of the electron-electron interaction term is underestimated by this pseudopotential. 

The pseudopotential in equation (\ref{eq:pt5}), on the other hand, yields comparable ionization potentials for all the atoms except helium. Part of the disparity between the Present$_2$ results and the reference values stem from the approximation in equation (\ref{eq:pt3}) and partly from the effect of the relativistic and other higher-order corrections which are expected to increase with the number of electrons.

\begin{table}[!ht]
    \centering
    \begin{tabular}{ccccc}
    \hline
    $S.No.$ & State & Present$_1$& Present$_2$& Ref.(eV)  \\
   \hline
   \hline
       $1$& 2s  & -5.500   & -4.977  & -5.390   \\                                                     
       $2$& 2p  & -5.500   & -3.967  & -3.542   \\ 
       $3$& 3s  & -2.444   & -2.033  & -2.016  \\                           
       $4$& 3p  & -2.444   & -1.760  & -1.555 \\ 
       $5$& 3d  & -2.444   & -1.748  & -1.511  \\ 
       $6$& 4s  & -1.375   & -1.099  & -1.048  \\                                  
       $7$& 4p  & -1.375   & -0.988  & -0.867  \\                                                                                   
       $8$& 4d  & -1.375   & -0.983  & -0.852  \\  
       $9$& 4f  & -1.375   & -0.983  & -0.848  \\                                 
    \hline
    \end{tabular}
    \caption{Numerically calculated eigenvalues for the excited lithium atom using the present pseudopotentials versus the reference values \cite{Angelfire2017}.  The present$_{1}$ and present$_2$ are the Hamiltonians given by equations (\ref{eq:pt1a}) and (\ref{eq:pt5}) as the pseudopotentials respectively. The results presented are truncated to $3$ d.p. }
    \label{tab2}
  \end{table}
In table \ref{tab2}, we present the excited state eigenvalues of lithium atom calculated using the two pseudopotentials in comparison with existing literature values \cite{Angelfire2017}. It is clear from the table that the pseudopotential in equation (\ref{eq:pt1a}) underestimates the effect of electron correlation for the excited states for any symmetry. It also shares the accidental degeneracy of the hydrogen-like atoms, differing with the experimental observations.
Equation (\ref{eq:pt5}) pseudopotential on the other hand yields excited state eigenvalues which are quite comparable with the reference data for all symmetries examined. As expected, the accidental degeneracy is removed in this case. In general, the second pseudopotential yields better results as compared to equation (\ref{eq:pt1a}), which is only excellent for spherically symmetric groundstate eigenvalues. 

\section{Conclusion}
We have derived two parameter-free pseudopotentials generalized for $n$-electron atoms. One of the pseudopotentials has an exact charge screening parameter which is invariant in space where as in the second case, the charge screening parameter is nearly-exact and varies in space. The two pseudopotentials are rather complementary. The first pseudopotential yields near-exact groundstate ionization potentials for helium, lithium, and berrylium atoms and unreliable results otherwise. On the other hand, the second pseudopotential yields reasonable groundstate ionization potentials for all atoms considered except helium.  Also, the excited state eigenvalues for lithium atom evaluated using the spatially varying screening parameter pseudopotential yieds results comparable to the literature data. The latter pseudopotential is therefore likely to be a candidate for further improvement in order to increase its validity with the experimental observations. The further improvement to the second pseudopotential, with varying charge screening parameter, can be considered by including the relativistic and higher-order corrections besides improving the approximation of the spatial distribution function.

\appendix
\section{} \label{app:A}
The potential energy function in equation (\ref{eq:pt1}) can be re-written as
\begin{equation}
V(r_i,r_j) = -\frac{Z}{r_i} + \sum_{j \neq i}^{n-1} \frac{r_i^2}{[r_i^2 + r_j^2]^{3/2}}
\end{equation}
where the exact partition function \cite{Jobunga2017b} for sharing the correlation energy is used.
We differentiate $V(r_i,r_j)$ with respect to $r_i$ and equate to zero to find the extremum values. Following the argument advanced in ref.\cite{Jobunga2017b} which ensures that the potential is minimized while treating the correlated electrons with an equal weight, we obtain the inequality
\begin{equation}
\frac{\partial V}{\partial r_i} = \frac{Z}{r_i^2} - \sum_{j \neq i}^{n-1} \frac{2\,r_i(r_i^2+r_j^2)}{[r_i^2+r_j^2]^{5/2}} \leq 0. \label{eq:appA2}
\end{equation}

We use the equality condition in equation (\ref{eq:appA2}) to obtain the correlation energy
\begin{equation}
\frac{1}{\sqrt{r_i^2 + r_j^2}} =\frac{\frac{Z\,f(r_i,r_j)}{2(n-1)}^{1/5}}{r_i}
\end{equation}
 between any two interacting electrons. 
\section{} \label{app:B}
The method through which the expectation value in equation (\ref{eq:pt3}) has been evaluated is shown in this appendix. The integral 
\begin{equation}
\begin{split}
 \langle f(r_i, r_j)^{\frac{3}{5}} \rangle =& \langle \phi (r_j)|\left[\frac{r_j^2}{r_i^2 +r_j^2}\right]^{\frac{3}{5}}|\phi(r_j)\rangle\\
           =& \int_{0}^{r_i} \mathrm{d}r_j\, \left[r_j^2\, t^{\frac{6}{5}}\left(1 + t^2\right)^{-\frac{3}{5}}\right]\exp (-2Zr_j)\\ & + \int_{r_i}^{\infty} \mathrm{d}r_j\,\left[r_j^2\, \left(1 + t^2\right)^{-\frac{3}{5}}\right]\exp(-2Zr_j) \label{eq:app1}
 \end{split}
 \end{equation}
 is evaluated in parts where we consider that ${0\leq r_j \leq r_i}$, ${r_i \leq r_j \leq \infty}$, ${t=r_</r_>}$, ${r_<=\mathrm{min}(r_i, r_j)}$, and ${r_>=\mathrm{max}(r_i, r_j)}$. We have used the hydrogenic orbital ${\phi(r_j) = \exp(-Zr_i)}$ for the probability density function and a binomial expansion
\begin{equation}
(1 + t^2)^{-\frac{3}{5}} = \sum_{k=0}^{\infty} \left(\begin{matrix} {-3/5}\\ k \end{matrix} \right)\, t^{2k}   \label{eq:app2}
\end{equation} 
  to evaluate the expectation value. Equation (\ref{eq:app1}) together with the series in equation (\ref{eq:app2}) yield an integral that cannot be evaluated exactly. In our case, only $k=0$ and $k=1$ are used for estimation. It is important to note that the expectation value in this case provides a static contribution of the correlated term to the active electron in the field of the other electron.
 

\bibliographystyle{apsrev}
\bibliography{/home/eric/Inworks/Literature}

\end{document}